\documentclass[aps,prl,twocolumn,amssymb, amsmath]{revtex4}
\usepackage{graphicx}
\usepackage{dcolumn}% Align table columns on decimal point 
\usepackage{bm}% bold math
\usepackage{natbib}

\newcommand{\EQ}[1]{Eq.~(\ref{eq:#1})}

\newcommand{\FIG}[1]{Fig.~\ref{fig:#1}}

\newcommand{\kT}{T}

\newcommand{\eps}{{\varepsilon_{\rm b}}}   % binding energy per basepair 
\newcommand{\G}{{B_{\ell}}}        % exponential loop cost function 
\newcommand{\F}{A}        % exponential loop cost function 
\newcommand{\Eini}{{\varepsilon_{\ell}}}  % loop initiation cost 
\newcommand{\cS}{{\cal S}}        % set of all basepairs in structure 
\newcommand{\cQ}{{\cal Q}}        % statisical weight of a basepairing pattern
\newcommand{\hW}{\hat{W}}    %z-transformed W 
\newcommand{\hZ}{\hat{Z}}    % z-transformed Z 
\newcommand{\hF}{\hat{A}}    % z-transformed A 
\newcommand{\hG}{\hat{B}}    % z-transformed F 
\newcommand{\xf}{{x^*}}     % critical y fugacity \newcommand{\yf}{{y^*}}     % critical y fugacity 
\newcommand{\dTc}{\tilde{T}_c}     % critical y fugacity
\newcommand{\sr}{\alpha}     % strand ratio
\newcommand{\hsr}{\bar{\alpha}}     % strand ratio
\newcommand{\hsrmax}{{\bar{\alpha}_{max}}}     % strand ratio

\begin{document}

\title{An intermediate  phase in  DNA melting}

\author{Richard A. Neher} 
\email{Richard.Neher@physik.lmu.de}

\author{Ulrich Gerland}
%\email{Ulrich.Gerland@physik.lmu.de}

\affiliation{Arnold Sommerfeld Center for Theoretical Physics 
and Center for Nanoscience (CeNS), LMU M\"unchen, Theresienstrasse 
37, 80333 M\"unchen, Germany}

\date{\today}

\begin{abstract} 
We predict a novel temperature-driven phase transition of DNA below the melting transition.
The additional, intermediate phase exists for repetitive sequences, when the two strands have 
different lengths. 
In this phase, the excess bases of the longer strand are completely absorbed as 
bulge loops inside the helical region. When the temperature is lowered, the 
excess bases desorb into overhanging ends, resulting in a contour length change. 
This continuous transition is in many aspects 
analogous to Bose Einstein condensation. 
Weak sequence disorder renders the transition discontinuous. 
\end{abstract}

\maketitle
The base-pairing interaction between the two strands of DNA is not only pivotal 
to its biological function \citep{Alberts:02}, but also leads to intriguing 
applications in nanotechnology \citep{Seeman:03}. One approach to probe this 
interaction is to monitor the DNA conformation as a function of temperature. 
Experimentally, one can observe the number of basepairs formed (using UV 
absorption \cite{Wartell:85,SantaLucia:04}), as well as changes of 
intra-molecular distances on the nanometer scale (using modern single-molecule 
techniques \cite{Zhuang:03}). On the theoretical side, the temperature 
dependence of DNA conformations has been studied for almost fifty years, using 
models of various degrees of complexity 
\cite{Zimm:59,Hill:59,Poland:66b,Peyrard:89,Cule:97,Kafri:00,Theodorakopoulos:00,Garel:04}. 
Particular attention has been paid to the characteristics of the melting 
transition, where the two strands separate completely. Whereas early models 
yielded only a crossover \cite{Zimm:59}, the Poland-Scheraga (PS) model 
\cite{Poland:66b} was the first to display a phase transition, albeit a 
continuous one, which appeared to be at variance with the experimentally 
observed sharp jump in the fraction of bound basepairs \cite{Wartell:85}. Only 
recently have mechanisms been proposed \cite{Kafri:00,Theodorakopoulos:00} 
which yield an abrupt, first order transition. 
So far, however, most analyses of DNA melting have incorporated only native 
interactions, i.e. base pairs that occur in the ground state of the molecule 
(see \cite{Hill:59,Garel:04} for notable exceptions). 
It is our aim here to show that such non-native interactions can introduce an 
intermediate phase in the melting behavior of DNA, associated with an additional 
conformational transition before strand separation.

Non-native interactions are particularly relevant for repetitive DNA sequences, 
which are common in genomes \citep{Lovett:04}. Periodic DNA, with e.g. a single 
base repeat such as \texttt{TTT\ldots} or a higher order repeat such as 
\texttt{CAGCAG\ldots}, can take on basepairing patterns with asymmetric loops 
and the two complementary strands can be shifted relative to each other. Here, 
we consider the general situation where the two strands can have arbitrary 
lengths $N$, $M$. We describe the DNA using a generalized PS 
model \cite{Garel:04} and calculate its equilibrium behavior analytically. 
We find that for $N\neq M$, the bound phase splits into two separate phases. 
The low temperature phase is characterized by an extensive length of the 
unbound end on the longer strand, whereas in the new intermediate phase these 
overhanging bases are absorbed into the helical region. Mathematically, and 
also conceptually, many aspects of this transition are analogous to 
Bose-Einstein condensation (BEC), as ``particles'' (bases) condense into a 
single ``state'' (the overhanging end), which thereby acquires macroscopic 
``occupation'' (length). Obviously, the analogy extends only to the behavior of 
the partition function, as there is no quantum coherence in the DNA problem. 
Effectively, the transition amounts to a temperature sensitive change in the 
contour length of the DNA molecule, which should be observable with optical or single 
molecule methods. While the transition is continuous for perfectly periodic 
sequences, we find that it becomes a first order transition once (weak) 
sequence disorder is introduced. We also show that the non-native interactions 
can change the order of the melting transition, as has been conjectured 
previously \cite{Kafri:02b}.
%MODEL FIGURE
\begin{figure}[b] \includegraphics[width=\columnwidth]{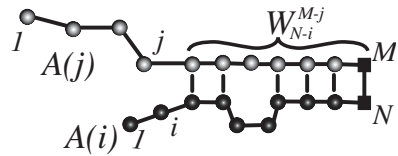}
\caption{\label{fig:melting_setup}
A possible configuration of two complementary DNA strands with 
a repetitive sequence (a bead represents one repeat unit). 
Note that repetitive sequences can form base pairing patterns 
with asymmetric loops. 
In general we allow for different strand lengths $N$, $M$. 
The last repeat units (squares) are permanently bound.  
}
\end{figure}

{\it DNA model.---}
We consider two DNA strands with lengths $N$ and $M\geq N$, respectively, and 
describe their interaction with a generalized PS model \cite{Garel:04,Neher:04}. 
Specifically, a base $i\le N$ of the lower strand can form a base pair 
$(i,j)$ with every complementary base $j\le M$ of the upper strand, whereas 
the formation of base pairs within a strand can be neglected (since 
we are interested only in sequences with a high degree of complementarity 
and a low degree of self-complementarity). 
Due to geometrical constraints, we may neglect the `crossing' of base 
pairs, e.g. two base pairs $(i_1,j_1)$ and $(i_2,j_2)$ with $i_1<i_2$ but 
$j_1>j_2$. 
The basepairing pattern $\cS$, i.e. the set of all formed base pairs, then 
creates a DNA conformation consisting of bound segments alternating with 
(possibly asymmetric) loops, see \FIG{melting_setup}. 
To simplify the discussion, we enforce the base pair $(N,M)$ at the right 
end, so that we need to consider only one overhanging end. 
Experimentally, this boundary condition would be realized e.g. by a 
few particularly strong basepairs at one end. 

To each basepairing pattern $\cS$, we assign a statistical weight $\cQ(\cS)$, 
which takes the form of a product with factors of four different types: 
(i) a Boltzmann factor $q=e^{\eps/k_BT}$ for every basepair with binding 
energy $-\eps<0$, 
(ii) a Boltzmann factor $g^2=e^{-\Eini/k_BT}$ for every loop with 
loop initiation cost $\Eini>0$, 
(iii) an entropic factor $\G(m)=s^m m^{-c}$ for each loop, which 
is the increase in the number of polymer configurations when 
$m$ bases form a (floppy) loop instead of being in a (rigid) double 
helical conformation, 
(iv) and a similar entropic factor $\F(n)=s^n n^{-\bar{c}}$ for a 
single-stranded end of $n$ bases. 
Here, the exponents $c,\bar{c}$ in the entropic factors are universal in 
that they are independent of the detailed polymer properties, but are 
sensitive to excluded volume interactions. 
For interacting self-avoiding loops one has $c\approx2.15$, while 
$\bar{c}\approx 0.1$ \citep{Kafri:02b}. 
Whereas the value of $c$ determines the critical behavior at the melting 
transition \cite{Kafri:00}, the non-universal constant $s$ has no 
qualitative effect on the melting behavior (we use $s=10$ in all 
numerical examples). 
In the following, we apply the DNA model to perfectly periodic 
sequences, where each repeat unit can be treated as an effective base 
with renormalized parameters (we use $\eps=6$ and 
$\Eini=3$ in temperature units, $k_B=1$). 
We emphasize that our simplistic model for the involved energies and 
entropies is meant to illustrate the physical phenomena in 
a transparent way, but leads 
to an unrealistic temperature scale. With a more detailed description  
\cite{SantaLucia:04}, we find that all of the interesting behavior happens at 
accessible temperatures \citep{future}.

{\it Free energy of periodic DNA.---}
To obtain the equilibrium properties of the DNA model, we calculate the 
partition sum over all basepairing patterns, $Z_N^M=\sum_{\cS}\cQ(\cS)$. 
By separating the single stranded ends from the double stranded 
part, see \FIG{melting_setup}, we write $Z_N^M$ as 
\begin{equation}
\label{eq:completeZ}
Z_N^M=\sum_{i=0}^{N-1}\sum_{j=0}^{M-1}\F(i)\F(j)\,W_{N-i}^{M-j}\;.
\end{equation}
Here, $W_{r}^t$ is the partition function of two complementary and periodic 
strands of length $r$ and $t$ with the first and last base pair formed. 
$W_{r}^t$ obeys the recursion relation 
\begin{equation}
\label{eq:recursion}
W_{r+1}^{t+1}=qW_r^t+g^2q\sum_{k+m>0}^{k<r,m<t}\G(k\!+\!m)W_{r-k}^{t-m},
\end{equation}
with the initial conditions $W_{1}^1=q$ and $W_{1}^{i}=W_{i}^1=0$ for 
$i>1$ \cite{Garel:04,Neher:04}. 
Eqs.~(\ref{eq:completeZ}) and (\ref{eq:recursion}) can be used to 
calculate $Z_N^M$ for finite lengths $N$, $M$. 
To extract the thermodynamic behavior in the limit of long strands, 
we take the $z$-transform $\hat{Z}(x,y)=\sum_{N,M=0}^\infty Z_N^M\,x^Ny^M$. 
Compared to the related case of a single self-complementary RNA strand 
folding back onto itself \citep{Bundschuh:99,deGennes:68}, we need two 
instead of one transformation variables here, due to the second strand of DNA. 
One obtains 
\begin{equation}
\label{eq:Z_melting}
\hZ(x,y)=\frac{\hF(x)\hF(y)qxy}{1-qxy+\frac{qg^2xy}{x-y}(y\hG(y)-x\hG(x))}\;,
\end{equation}
where the transforms of the entropic factors are given by 
$\hF(z)=\phi_{\bar{c}}(sz)+1$ and $\hG(z)=\phi_c(sz)$, with the 
polylogarithm $\phi_c(z)=\sum_{n=1}^\infty z^n n^{-c}$. 

\begin{figure}	
\includegraphics[width=0.55\columnwidth]{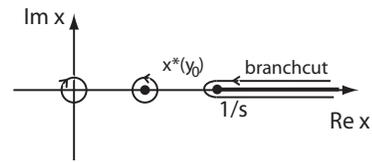}
\caption{\label{fig:contourint} 
The contour integration required for the inverse transformation of $\hZ(x,y_0)$ is given by 
the sum of the integral along the branchcut $[s^{-1},\infty]\subset\mathbb{R}$ 
and the integral encircling the singularity 
of $\hW(x,y_0)$ at $x=\xf(y_0)$. This isolated singularity only exists 
below the melting temperature. 
}
\end{figure}

The $z$-transformation carried out above amounts to a change from the canonical 
to the grand canonical ensemble. The transformation variables $x,y$ play the 
role of fugacities for bases in the lower and upper strand, respectively. 
However, for the ensuing discussion, it is advantageous to keep the length $N$ 
of the shorter strand fixed as a reference. 
Hence, we perform the inverse transformation for the lower strand by contour 
integration in $x$, see \FIG{contourint}, to obtain the partition sum 
$Z_N(y_0)$ for $N$ bases on the lower strand and the upper strand coupled to a 
``nucleotide reservoir'' with fixed fugacity $y_0$. 
Whenever both strands are bound, $\hZ(x,y)$ has a singularity at 
$\xf(y_0)<s^{-1}$, see \FIG{contourint}. For large $N$, the contour integration 
is dominated by the residue at $\xf(y_0)$, leading to 
$Z_N(y_0)=\hF(y_0)\xf(y_0)^{-N}$. Hence, the free energy of the bound phase 
is given by $N f_b(y_0)-\kT\ln\hF(y_0)$, where the first term is the 
contribution of the helical region with a free energy per length 
$f_b(y_0)=\kT \ln\xf(y_0)$, and the second term is the contribution from the 
unbound end of the longer strand.
%FIGURE COMPLEX PLANE
The free energy for given $N$ and $M$ is then obtained by saddle point 
integration, 
\begin{equation}
\label{eq:freeenergy}
\frac{F(T,N,M)}{\kT}=-\ln\hF(y_0)+ N \frac{f_b(y_0)}{\kT}+M\ln(y_0),
\end{equation}
where the fugacity $y_0$ is determined by
\begin{equation}
\label{eq:meanlength}
\begin{split}
M=\langle M \rangle_{y_0}&= y_0	\frac{\partial \ln\hF(y_0)}{\partial 
y_0}-N\frac{y_0}{\kT}\frac{\partial f_b(y_0)}{\partial y_0}\;.
\end{split}
\end{equation}

%BEC AND MELTING FIGURE 
\begin{figure}[bt] \includegraphics[width=0.9\columnwidth]{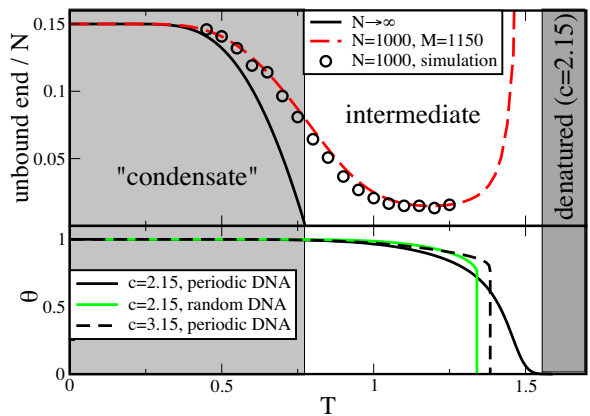}
\caption{\label{fig:condensate}
%The two DNA observables as a function of temperature.
Top: The length of the unbound end, normalized by the total length $N$ of the 
shorter strand. For finite systems ($N=1000$, dashed line), the unbound end shrinks 
to a minimal value and increases again, as the melting temperature is 
approached. Monte Carlo simulation data (circles) agrees well 
with the analytical result. In the $N\to\infty$ limit, the overhang length diverges below 
$T_c$ and is of order 1 for $T>T_c$. 
Bottom: The fraction of bound basepairs $\theta$ as a function of 
temperature. For periodic sequences with $c=2.15$, $\theta$ vanishes with 
zero slope, whereas a random sequence shows a first 
order phase transition. When increasing $c$ to 3.15, the periodic sequence 
displays a similar first order transition.	 }
\end{figure}

{\it Phase diagram.---}
To extract the physical behavior of the DNA model from Eqs.~(\ref{eq:freeenergy}) 
and (\ref{eq:meanlength}), we focus on two observables, the total number of 
base pairs, $N\theta$, and the length of the single-stranded overhang. 
The fraction $\theta$ of bound base pairs is calculated from the free 
energy per length of the helical region as 
\begin{equation}
\label{eq:theta}
\theta= -\frac{q}{\kT} \frac{\partial f_b(y_0)}{\partial q}\;.
\end{equation}
To obtain the overhang length, we note that the right hand side of 
(\ref{eq:meanlength}) decomposes the total length $M$ of the upper 
strand into two contributions, where the first term is the 
expected overhang length and the second term corresponds to 
the number of bases in the helical region. 
The dashed line in \FIG{condensate} (top) shows the overhang length 
as a function of temperature, for $N=1000$ and $M=1150$. 
At low temperatures, the two DNA strands are completely aligned, so 
that all $M-N$ excess bases of the longer strand form an overhanging 
end. However, we observe that the overhang length decreases with 
increasing temperature, dropping almost to zero before it rises again 
sharply at even higher temperature. 
We see in \FIG{condensate} (bottom) that this drop occurs in a 
temperature range where almost all possible base pairs are formed, 
and the rise occurs when the two strands separate. 
These observations suggest that a temperature-driven conformational 
transition occurs before the melting transition. 

\begin{figure}[tb] 
\includegraphics[width=0.5\columnwidth]{fig4a}
\includegraphics[width=0.44\columnwidth]{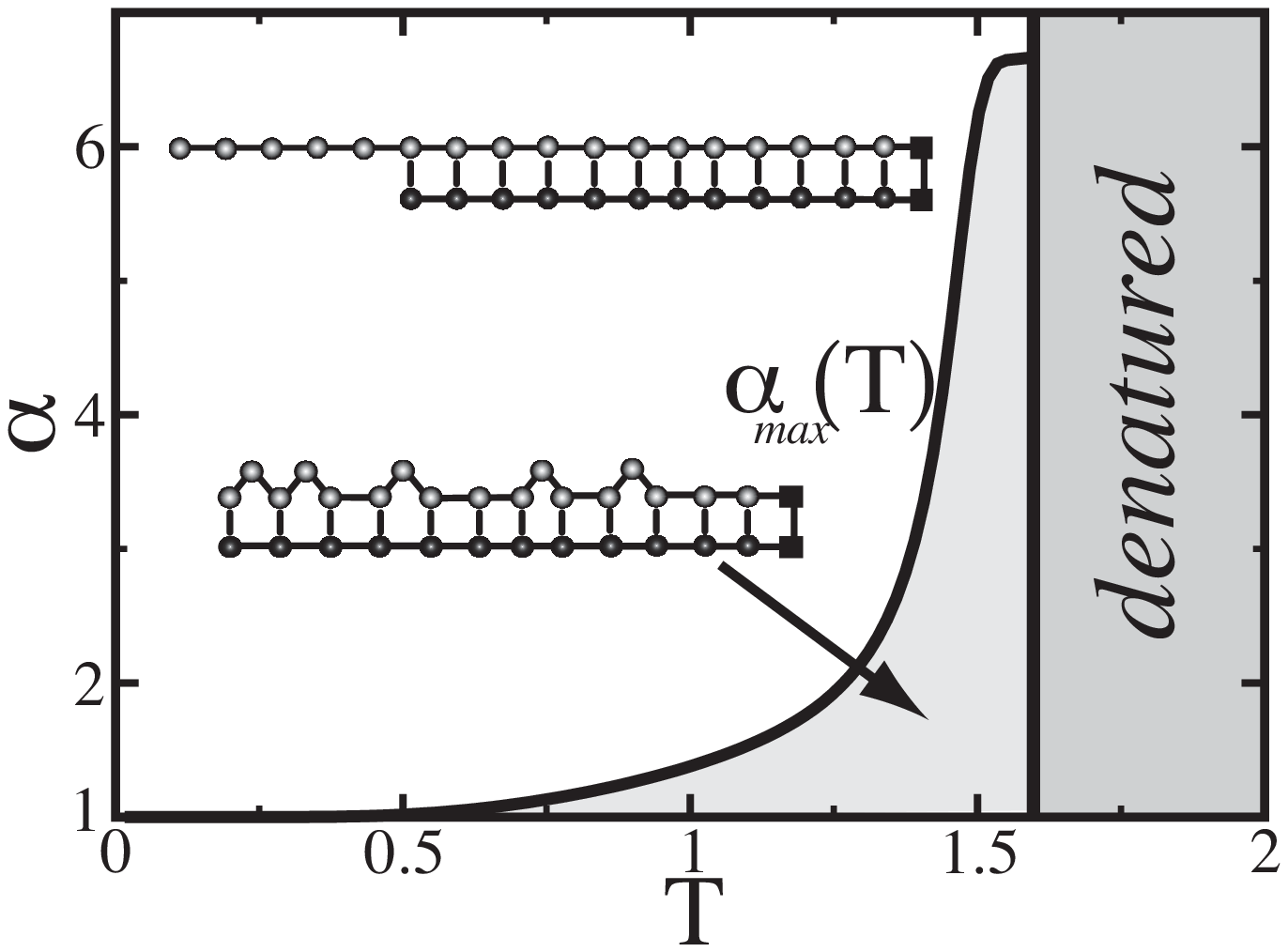} 
\caption{\label{fig:bec} 
Left: The fugacity $y_0$ 
vs. $T$ for different system sizes $N$. In the thermodynamic limit, 
$y_0=s^{-1}$ for $T<T_c$. As for BEC, $y_0$ approaches its limiting value as $s^{-1}-y_0\sim 1/N$.
Right: Phase diagram of periodic DNA. At low temperatures, both strands are 
completely aligned and excess bases of the longer strand form an unbound end. 
In the intermediate phase, all excess bases are absorbed into the helical region. 
}
\end{figure}

This transition is in fact completely analogous to BEC, as 
\EQ{meanlength} parallels the behavior of the 
equation of state for an ideal Bose gas: 
If we divide \EQ{meanlength} by our system size $N$ and introduce the 
``particle density'' $\alpha=M/N$, we obtain  
\begin{equation}
\label{eq:alpha}
\alpha=\frac{1}{N}\,\frac{\phi_{\bar{c}-1}(s\,y_0)}{\phi_{\bar{c}}(s\,y_0)+1}+\hsr(y_0) \;,
\end{equation}
where $\hsr(y_0)=-\frac{y_0}{\kT}\frac{\partial f_b(y_0)}{\partial y_0}\geq 1$
is the density inside the helical region. 
In \EQ{alpha}, the first term on the right hand side corresponds to the occupation 
of the ground state of an ideal Bose gas, whereas $\hsr(y_0)$ is analogous 
to the occupation of the excited states. 
The fugacities of a Bose gas and our DNA are bounded: for the former, by the 
energy of the ground state, and for the DNA by the weight of an unbound monomer, i.e. $y_0\le s^{-1}$.  
The population of the excited states increases monotonically with the fugacity, 
and attains a finite maximal value, in our case $\hsrmax=\hsr(s^{-1})$ 
(provided the loop exponent $c>2$). 
If the density $\alpha$ exceeds this maximal value
the length of the unbound end has to diverge in 
order to accommodate the remaining bases. In other words, the length 
of the unbound end becomes extensive. In an analogous way, the ground state of 
a Bose gas is macroscopically populated at low temperatures.
In this ``condensate'' phase, the fugacity is locked to the value $s^{-1}$ in 
the thermodynamic limit ($N,M\rightarrow\infty$, $\sr=const.$). 
The deviation for finite systems scales as $s^{-1}-y_0\sim 1/N$, see 
\FIG{bec} (left). 
In the opposite case, where $\alpha<\hsrmax$, there is a solution to \EQ{alpha}
with $y_0<s^{-1}$ and the unbound end remains finite for all system sizes. 

%FUGACITY FIGURE

It is easily shown that $\hsrmax$ approaches 1 at low temperatures, and 
consequently all excess bases of the longer strand are condensed in the 
overhang, as illustrated in \FIG{bec} (right). 
As $T$ increases, more and more bases are absorbed in the helical region ($\hsrmax$ increases),
and the system enters the intermediate phase at $T=T_c$, where $\hsrmax=\sr$.
At $T_c$ the condensate fraction vanishes, 
 as the solid line shows in \FIG{condensate} (top). 
If $T$ is raised to the melting temperature $T_m$, which is 
independent of $\sr$, the strands separate and $\theta$ vanishes 
(denatured phase). Note that the intermediate phase exists only when $\sr$ is not 
too large. 

It has been previously predicted \cite{Kafri:02b} that the loop 
exponent $c$ is effectively reduced by one for periodic sequences 
compared to the standard PS-model with native base pairs only. 
This prediction is explicitly confirmed by our exact calculation of 
the free energy. 
We find \cite{future}, that there is no melting transition if $c\leq 2$, that 
the transition is continuous if $2< c \leq3$ and of first order if $c>3$. 
For $2< c\leq 3$, we obtain $\theta\sim |T-T_m|^{\frac{3-c}{c-2}}$, using the 
same method as \cite{Fisher:66} for the standard PS-model. 
To illustrate this, we plot $\theta$ for periodic sequences and for 
the standard PS-model in \FIG{condensate} (bottom). Whereas for the latter 
$\theta$ drops discontinuously to zero for $c=2.15$, $\theta$ of periodic DNA 
vanishes with zero slope. Only after increasing $c$ artificially to $3.15$ 
does periodic DNA exhibits a similar first order transition. 

%DISORDER FIGURE
\begin{figure}[tb]
\includegraphics[width=0.47\columnwidth]{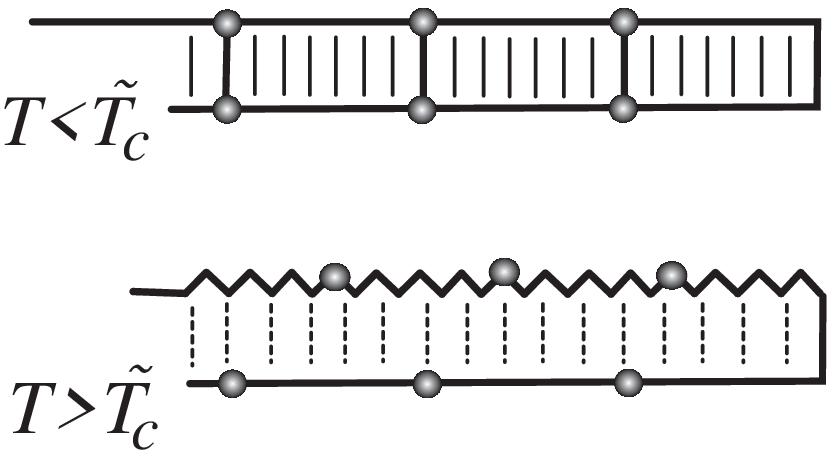} 
\includegraphics[width=0.47\columnwidth]{fig5b}
\caption{\label{fig:mutations}
Mutations in the sequence renders the transition to the intermediate phase 
discontinuous. The plot shows Monte Carlo data for the length of the unbound end for 
evenly spaced mutations every 100 and 200 bases. This length  
drops discontinuously at a critical temperature $\dTc$, which depends on the 
mutation density and the binding strength of mutated basepairs.}
\end{figure}

{\it Weak sequence disorder.---}
Is the intermediate phase identified above robust against sequence disorder? 
To address this question, we replace a small fraction of base pairs by 
bases that can pair with each other, but not with other bases in 
the sequence. 
Fig.~5 shows the average length of the overhanging end, obtained by Monte 
Carlo simulation, for evenly spaced mutations with densities $0.005$ and $0.01$ and mutation 
strength $\bar{\varepsilon}_b=2$ . The plot suggests that in the presence of 
weak sequence disorder the transition described above remains, but is of 
first order instead of being continuous. 
The unbound end keeps its ground state length up to certain 
temperature, and then shortens rapidly.
This is readily understood, when comparing the energy barriers for forming 
bulgeloops with and without mutations. The formation of a bulgeloop on the 
longer strand of a perfectly periodic molecule requires only the initiation 
energy $\Eini$. In the presence of mutation, however, shifting both strands 
breaks mutated basepairs. Hence, to form a bulge loop, all mutations to the 
left of the loop have to be broken and the energy barrier for loop formation 
grows with the distance from the end. Due to this extensive energy barrier for 
loop formation, mutated basepairs stay bound in a finite temperature range. For 
a sufficiently low density of mutations, there is a temperature $\dTc$, at 
which the entropy gained by distributing excess bases in loops along the 
molecule outweighs the energetic costs \citep{Neher:05}. Below $\dTc$ all 
mutations are bound, if $T>\dTc$ as many mutations open, as are necessary to 
absorb all excess bases. On increasing the mutation density, $\dTc$ 
approaches the melting temperature and the intermediate phase vanishes.

{\it Discussion.---}
We have identified a BEC-like conformational transition in periodic and nearly 
periodic DNA, which occurs below the melting transition. 
This transition leads to a change in the contour length of the DNA molecule, 
which is roughly proportional to $M-N$. 
We also expect an effect on the persistence length of the helical region due 
to the increased density of bulgeloops. 
The hallmark of the transition, i.e. the shortening of the unbound end, could be 
directly observed by resonant energy transfer between fluorescent dyes 
located at the ends of the two strands. We expect the 
existence of the intermediate phase to be independent of the details of our model. 
Furthermore, we have shown that the additional conformations possible for repetitive sequences change 
the critical behavior at the melting transition. 

We are grateful for important comments by E.~Frey and H.~Wagner. 
We acknowledge financial support by the {\it Deutsche 
Forschungsgemeinschaft} through the Emmy Noether Program.

\end{document}